\documentclass[manuscript,amsmath,amssymb,superscriptaddress]{revtex4-2}

\usepackage{tabularx}
\usepackage{bm}
\usepackage{graphicx}

\usepackage{hyperref}
\hypersetup{colorlinks=true,urlcolor= blue,citecolor=blue,linkcolor= blue,bookmarks=true,bookmarksopen=false}

\usepackage{color}

\usepackage{amsmath,mathtools}
\usepackage{multirow}
\usepackage{dcolumn}
\usepackage{amssymb,amscd,xypic,bm,wasysym}
\usepackage{float}
\usepackage{cleveref}
\usepackage[caption=false,position=top,captionskip=0pt,farskip=0pt]{subfig}
\usepackage{soul}

\begin{document}
	
\title{Current-induced quasiparticle magnetic multipole moments}
	
\author{Muhammad Tahir}
\affiliation{Department of Physics, Colorado State University, Fort Collins, CO 80523, USA}
\author{Hua Chen}
\affiliation{Department of Physics, Colorado State University, Fort Collins, CO 80523, USA}
\affiliation{School of Advanced Materials Discovery, Colorado State University, Fort Collins, CO 80523, USA}

\begin{abstract}
Magnetic ordering beyond the standard dipolar order has attracted significant attention in recent years, but it remains an open question how to effectively manipulate such nontrivial order parameters using external perturbations. In this context, we present a theory for Cartesian magnetic multipole moments and their currents created by electric currents based on a general gauge-invariant formula for arbitrary-order spin magnetic multipole moments of Bloch wave packets. As a concrete example, we point out that the low-energy quasiparticles in phosphorene subject to a perpendicular electric field have a valley structure that hosts magnetic octupole moments. The quasiparticle magnetic octupole moments can be exhibited by an in-plane electric current and lead to accumulation of staggered spin densities at the corners of a rectangular-shaped sample. A current carrying the octupole moments can further be induced through nonlinear response. Our work paves the way to systematically searching for and utilizing quasiparticles with higher-order magnetic multipole moments in crystal materials.
\end{abstract}

\maketitle

\section{Introduction}\label{sec:intro}
Magnetic order that goes beyond the conventional collinear ferro-, ferri- and antiferro-magnetic ordering has been studied in-depth historically \cite{Andreev1978,  Andreev_1980, DZYALOSHINSKII1992579, Andreev1996, Astrov1996}, and got revived interest in recent years due to new experimental and theoretical advances that allow convenient characterization and manipulation of the corresponding order parameters, which are often described by higher-order electromagnetic multipole moments \cite{kuramoto_2009,santini_2009,sakai_2011,onimaru_2011,onimaru_2016,kubo_2004,mannix_2005,kuwahara_2007,matsumura_2009,watanabe_2018,hayami_2018,gao_2018,gao_2018_2,shitade_2018,shitade_2019,Dubovik_1990,Gorbatsevich_1994,ederer_2007,Spaldin_2008,arima_2005,VanAken_2007,Hayami_2014,Batista_2008,thole_2016,watanabe_2017_1,Suzuki2017}. Since multipolar order parameters by symmetry do not couple to uniform dipolar fields, it has been challenging both to characterize and to manipulate them. For certain collinear or noncollinear antiferromagnets whose magnetic order parameters transform similarly as the magnetic dipole, it has been established in recent years that they can be measured using experimental setups designed for ferromagnets, e.g. the anomalous Hall effect, anomalous Nernst effect, magneto-optical Kerr effect, etc., and be manipulated using uniform magnetic fields coupled through the symmetry-allowed weak net magnetization \cite{Solovyev1997, Tomizawa2009, Ohgushi2000, Shindou2001,chen_2014, Kubler_2014, Nakatsuji_2015, Nayak2016, Zhou2019, Gurung2019, Boldrin2019, Zhao2019, Liu2018, Smejkal2020, chen_2020, chen_2022}. However, it is more desirable to have external perturbations that directly couple to the multipolar order parameters by symmetry. On the other hand, the rapid development of the field of spintronics in the last decade has suggested that quasiparticle currents carrying the appropriate quantum number--spin, even if it is not exactly conserved, can be highly efficient in manipulating ferromagnetic magnetization \cite{wolf_2001,RALPH20081190,bader_2010,berger_1996,SLONCZEWSKI1996L1,Brataas2012,manchon_2019}. It is therefore interesting to ask if quasiparticle currents carrying higher-order multipole moments can be harnessed for effectively controlling multipolar magnetic order.

However, an immediate difficulty arises when trying to generalize the concept of quasiparticle spin currents to that of multipole moments. Namely, multipole moments as quantum numbers of quasiparticles are not as well defined as the electron spin. In particular, even the macroscopic multipole moments of equilibrium condensed matter systems have certain ambiguities due to the nonvanishing boundary contribution in the thermodynamic limit. A famous example is the electric polarization \cite{Resta_1994,king-smith_1993}, the value of which in a macroscopic system can be changed by moving charges between opposite surfaces even if the bulk part of the system's Hamiltonian is unchanged. In the modern theory of electric polarization such an indeterminacy is rooted in the gauge dependence of the Berry connection of Bloch electrons. Through the pioneering studies on the electric polarization and orbital magnetization \cite{xiao_2005,thonhauser_2005,shi_2007}, it is now understood that gauge-invariant expressions of similar multipole moments that involve only the bulk Bloch states, such as the spin and orbital toroidization \cite{gao_2018,gao_2018_2,shitade_2018,shitade_2019} and the electric quadrupole moment \cite{lapa_2019,daido_2020}, can be obtained through thermodynamic considerations. Namely, they are defined as the thermodynamic variables conjugate to the appropriate spatial gradients of electromagnetic fields. But the procedure for obtaining the gauge-invariant expressions is nontrivial and general results of higher-order multipole moments have not been given in literature. 

Separately, physical observables that are vanishing in the equilibrium state of a system may become nonzero in nonequilibrium. For example, a nonmagnetic heavy metal can host nonequilibrium spin accumulation and spin currents, as long as the essential symmetry requirements for the existence of these observables are satisfied due to the external forces responsible for the nonequilibrium. It is therefore reasonable to expect that systems that lack an equilibrium magnetic multipole order may still generate nonequilibrium multipole moments or their currents. To describe the creation of such nonequilibrium quantities, a physically intuitive perspective is to consider quasiparticles that individually carry magnetic multipole moments. For example, both the spin polarization and spin currents can be understood as due to the imbalance between quasiparticles with different spins and/or spin current densities in nonequilibrium. In particular, it is instructive to look for materials that have valleys in their momentum space that host approximately momentum-independent multipole moments, so that a selective creation of quasiparticles in such valleys and their currents directly lead to the corresponding multipolar observables.  

In this paper we provide a framework for defining and calculating arbitrary-order Cartesian magnetic multipole moments of Bloch wave packets, which is analogous to the wave-packet understanding of spin magnetic moment of Dirac electrons \cite{huang_1952,Chang_2008}. In Sec.~\ref{sec:mdef}, by starting from a wave packet construction, we identified the common structure of the spin magnetic multipole moments that are independent of the wave packet shape and are also gauge invariant. We then derived a general gauge-invariant formula for arbitrary-order spin magnetic multipole moments. As a concrete example, in Sec.~\ref{sec:bp} we showed that the low-energy Hamiltonian of phosphorene in the presence of a perpendicular electric field has a valley structure and hosts spin octupole moments near the Brillouin zone center. Such multipole moments can then be brought into real space by applying an in-plane transport current, and consequently lead to staggered spin accumulations at the corners of a square sample and to octupole currents through a nonlinear response, as discussed in Sec.~\ref{sec:current}. Finally we give a brief discussion and outlook in Sec.~\ref{sec:discussion}.

\section{Spin magnetic multipole moments of Bloch wave packets}\label{sec:mdef}
A wave packet of a non-degenerate Bloch band labeled by $n$ at momentum $\mathbf k_c$ is defined as $| W \rangle = \int_{\rm BZ} d^3\mathbf k w_{\mathbf k} |n\mathbf k\rangle$ where $w_{\mathbf k}$ is a scalar function localized at $\mathbf k= \mathbf k_c$. The case of degenerate bands will be discussed in a future work. A naive definition of the Cartesian $2^l$-order spin magnetic multipole moment of the wave packet is 
\begin{eqnarray}\label{eq:Mg}
	(\mathcal{M}^g)^{i_l}_{i_1 i_2\dots i_{l-1}} \equiv \langle W | \prod_{n=1}^{l-1} r_{i_n} s_{i_l} | W\rangle
\end{eqnarray}
which, however, clearly depends on the choice of origin. This can already be seen by considering the center-of-mass position of the wave packet $\mathbf r_c$, which also has the meaning of its (electric) dipole moment:
\begin{eqnarray}
\mathbf r_c = \frac{(2\pi)^3}{V_{\rm uc}}\int_{\rm BZ} d^3\mathbf k \left[i(w_{\mathbf k}^*\partial_{\mathbf k} w_{\mathbf k})  + \langle u_{n \mathbf k} | i\partial_{\mathbf k} | u_{n \mathbf k}\rangle |w_{\mathbf k}|^2 \right]
\end{eqnarray}
where $|u_{n \mathbf k}\rangle$ is the periodic part of the Bloch state $|n \mathbf k\rangle $. Due to the $\partial_{\mathbf k}$ in $\mathbf r_c$ the latter is not invariant under a $\mathbf k$-dependent $U(1)$ gauge transformation on either $w_{\mathbf k}$ or $|u_{n \mathbf k}\rangle$, which reflects the indeterminacy of the absolute position of a wave packet. As a well-known result in classical electromagnetism, only the lowest-order nonvanishing multipole moment of a given charge or current distribution is origin independent \cite{Jackson:490457}. Since the wave packet is normalized, its dipole moment must be position dependent. Nonetheless, by considering the shift of the center-of-mass position of the wave packet under a continuous variation of certain parameters that the Hamiltonian depends on, one can obtain a gauge-invariant result, which underlies the modern theory of polarization \cite{king-smith_1993,xiao_2005}. 

Separately, the localized nature of a wave packet makes it sensible to define multipole moments by using the center-of-mass as the origin. Physically illuminating results have been obtained by this definition, most prominently the spin magnetic moment of a Dirac electron wave packet \cite{huang_1952,Chang_2008}, and more recently the orbital magnetic moment \cite{xiao_2005,thonhauser_2005,shi_2007} and the magnetic toroidal moment \cite{gao_2018,gao_2018_2} of Bloch wave packets. Moreover, these results are also gauge-invariant, making it sensible to discuss the transport of such observables by the center-of-mass motion of the wave packets. 

In the same spirit, one may be tempted to define the $2^l$-order spin magnetic multipole moment of a Bloch wave packet by choosing its center-of-mass as the origin. However, different from the low-order ($l\le 3$) multipole moments, at higher orders such a naive construction does not in general lead to a gauge-invariant expression. Instead, we define a general $2^l$ magnetic multipole moment of a Bloch quasiparticle as
\begin{eqnarray}\label{eq:Mw}
\mathcal{M}_{j_1 j_2\dots j_{l-1}}^{j_l} \equiv \mathcal{N}^{-1}   {\rm Re} \sum_{\{j\}_{l-1}^{\rm u}} \langle u_{n \mathbf k} | s_{j_l} \prod_{j\in \{j\}_{l-1}^{\rm u}} (i\partial_{k_j} - \mathcal{A}_j) | u_{n \mathbf k}\rangle \big |_{\mathbf k = \mathbf k_c}
\end{eqnarray}
which is both gauge-invariant and independent of the shape of the wave packet $w_{\mathbf k}$. Here the Berry connection $\mathcal{A}_j \equiv \langle u_{n \mathbf k} | i\partial_{k_j} | u_{n \mathbf k}\rangle $; the summation is over all non-repeated permutations of $l-1$ elements $\{j_1, \dots, j_{l-1}\}$. The normalization factor $\mathcal{N} = (l-1)!/(N_x! N_y! N_z!)$ where $N_j$ is the number of times that the Cartesian index $j$ appears in the set $\{j_1, \dots, j_{l-1}\}$. The summation over $\{j\}_{l-1}^{\rm u}$ ensures that $\mathcal{M} $ is always invariant under arbitrary permutations of its spatial components.

Eq.~\eqref{eq:Mw} is the first major result of this work. Details of the derivation leading to it are included in Sec.~I of \cite{supp} and we sketch the main idea here. Our starting point is the following identity:
\begin{eqnarray}\label{eq:ginv1var}
(i\partial_x - A - i\partial_x \ln |g| + \partial_x \arg g)^n (gf) = g (i\partial_x - A)^n f
\end{eqnarray}
where $n>0$ is an arbitrary integer, $A, g, f$ are arbitrary functions of the variable $x$, with $g$ additionally required to have nonvanishing norm and smooth argument across potential branch cuts. 

\emph{Proof.} For $n=1$ Eq.~\eqref{eq:ginv1var} obviously holds. Assume it holds for a given integer $k\ge 1$, it follows that
\begin{eqnarray}
&&\left(i\partial_x - A - i\partial_x \ln |g| + \partial_x \arg g \right)^{k+1} (gf) \\\nonumber
&=& \left(i\partial_x - A - i\partial_x \ln |g| + \partial_x \arg g \right) \left[g \left(i\partial_x - A \right)^{k} f\right]\\\nonumber
&=& g \left(i\partial_x - A \right)^{k+1} f
\end{eqnarray}
Therefore Eq.~\eqref{eq:ginv1var} holds for any $n >0$. QED

Eq.~\eqref{eq:ginv1var} can also be generalized to the case of multivariable functions \cite{supp}. To apply it we convert Eq.~\eqref{eq:Mg} to
\begin{eqnarray}
	\mathcal{M}^g = \frac{(2\pi)^3}{V_{\rm uc}} \int d^3\mathbf k w_{\mathbf k} \langle u_{n \mathbf k} | \mathbf s (i\partial_{k_x})^{N_x}  (i\partial_{k_y})^{N_y}  (i\partial_{k_z})^{N_z}  (w_{\mathbf k} |u_{n \mathbf k}\rangle )
\end{eqnarray}
Eq.~\eqref{eq:ginv1var} further motivates us to substitute $i\partial_{k_j}$ by $i\partial_{k_j} -\mathcal{A}_j -i\partial_{k_j} \ln |w_{\mathbf k}| + \partial_{k_j} \arg w_{\mathbf k} $ in the last result. This replacement leads to an expression that is both gauge invariant and independent of the shape of the wave packet (up to a prefactor as discussed below). Transformation for the shape of the wave packet is defined as $w_{\mathbf k}\rightarrow g_{\mathbf k} w_{\mathbf k}$ as long as $g_{\mathbf k}$ and $w_{\mathbf k}$ satisfy the requirements on $g$ mentioned above, and both $w_{\mathbf k}$ and $g_{\mathbf k}w_{\mathbf k}$ are peaked at $\mathbf k_c$. It then follows from Eq.~\eqref{eq:ginv1var} that such a transformation simply amounts to $|w_{\mathbf k}|^2 \rightarrow |g_{\mathbf k} w_{\mathbf k}|^2$ in the integrand which does not change $\mathbf k_c$. The integral weighted by $|g_{\mathbf k} w_{\mathbf k}|^2$ is then approximated by the integrand evaluated at $\mathbf k_c$. Finally to recover the invariance of $\mathcal{M}$ under permutations of its Cartesian indices a symmetrization is introduced through the summation in Eq.~\eqref{eq:Mw}. 

In Sec.~II of \cite{supp} we showed that Eq.~\eqref{eq:Mw} reproduce the wave-packet contributions in the previous results of spin quadrupole or toroidal moments. As new formulas resulting from Eq.~\eqref{eq:Mw} we give below those for the spin octupole ($l=3$) and hexadecapole ($l=4$) \cite{watanabe_2017_1} moments:
\begin{eqnarray}\label{eq:l34formula}
	\mathcal{M}_{ab}^c &=& -\langle s_c \partial_a\partial_b \rangle \big |_{\boldsymbol{\mathcal{A}} = 0}\\\nonumber
	\mathcal{M}_{abc}^d &=& -\frac{1}{6}\langle i s_d\partial_a\partial_b\partial_c \rangle  \big |_{\boldsymbol{\mathcal{A}} = 0} + \frac{1}{6}\partial_a\partial_b \mathcal{A}_c\langle s_d\rangle \big |_{\boldsymbol{\mathcal{A}} = 0}  \\\nonumber
&&+ \frac{1}{6}[\partial_a\mathcal{A}_b \langle s_d \partial_c\rangle + \partial_b\mathcal{A}_c \langle s_d \partial_a\rangle + \partial_a\mathcal{A}_c \langle s_d \partial_b\rangle ] \big |_{\boldsymbol{\mathcal{A}}= 0} + (a,b,c\,{\rm permutations})
\end{eqnarray}
where $\langle\dots\rangle \equiv {\rm Re} \langle u_{n \mathbf k} |\dots | u_{n \mathbf k}\rangle$. It is important to note that $\boldsymbol{\mathcal{A}} = 0$ does not necessarily mean that the derivatives of $\boldsymbol{\mathcal{A}}$ also vanish. In practice, when a finite-difference evaluation of the $k-$derivatives which requires a smooth gauge choice is not easily implemented, Eq.~\eqref{eq:l34formula} needs to be supplemented with explicit formulas of the $k-$derivatives of $|u_{n \mathbf k}\rangle$ in terms of known matrix elements of the Bloch Hamiltonian and its derivatives at $\mathbf k$, which we give in Sec.~II of \cite{supp}. In particular, the octupole moment thus obtained is
\begin{eqnarray}
	\mathcal{M}^c_{ab} &=&{\rm Re}\sum_{m\neq n} \frac{(s_c)_{nn}(\partial_a H_{\mathbf k})_{nm} (\partial_b H_{\mathbf k})_{mn} }{(\epsilon_{n \mathbf k} - \epsilon_{m \mathbf k})^2}-{\rm Re} \sum_{m\neq n} \frac{(s_c)_{nm}(\partial_a\partial_b H_{\mathbf k} )_{mn}}{\epsilon_{n \mathbf k} - \epsilon_{m \mathbf k}}\\\nonumber
&&-{\rm Re} \sum_{l\neq n,m\neq n}\left\{ \frac{(s_c)_{nm}(\partial_b H_{\mathbf k})_{ln}  (\partial_{a} H_{\mathbf k})_{ml}  }{(\epsilon_{n \mathbf k} - \epsilon_{l\mathbf k})(\epsilon_{n \mathbf k} - \epsilon_{m \mathbf k}) } +(a\leftrightarrow b)\right\} \\\nonumber
&&+ {\rm Re} \sum_{m\neq n}\left\{ \frac{(s_c)_{nm}(\partial_b H_{\mathbf k})_{mn} (\partial_a H_{\mathbf k})_{nn} }{(\epsilon_{n \mathbf k} - \epsilon_{m \mathbf k})^2 } +(a\leftrightarrow b)\right\}
\end{eqnarray} 
which will be used in the next section. Eq.~\eqref{eq:l34formula} together with their counterparts expressed in terms of such matrix elements is the second major result of this paper.

Realistically interesting cases that we consider are valleys in momentum space at which the multipole moments of the quasiparticles are approximately constant, similar to the orbital moments in monolayer 2H transition-metal dichalcogenides \cite{yao_2008,Zeng2012,Sie2015}. The time-reversal symmetry is broken as long as the valley is not at a time-reversal invariant high-symmetry point, even if the system in equilibrium has time-reversal symmetry. Moreover, the little group at the valley center must be compatible with the multipole moment under consideration. In the next section we use these guidelines to give an example of the spin octupole moments in phosphorene subject to a perpendicular electric field.

\section{Spin octupole moments in phosphorene}\label{sec:bp}
\begin{figure}[ht]
	\centering
	\subfloat[]{\includegraphics[width=2.6 in]{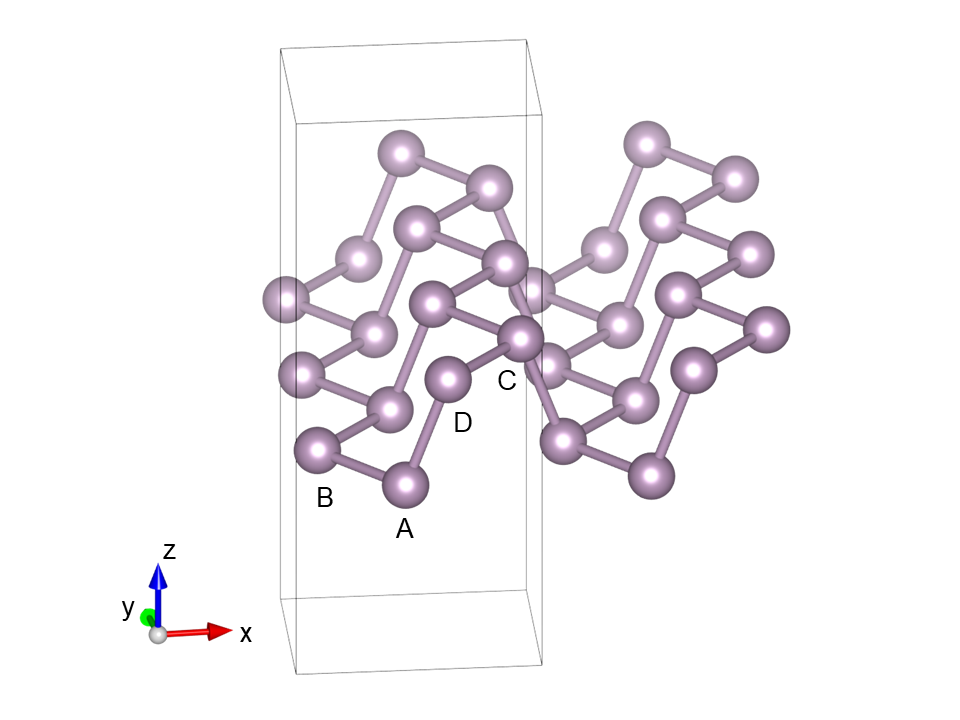}}\;\;
	\subfloat[]{\includegraphics[width=2.9 in]{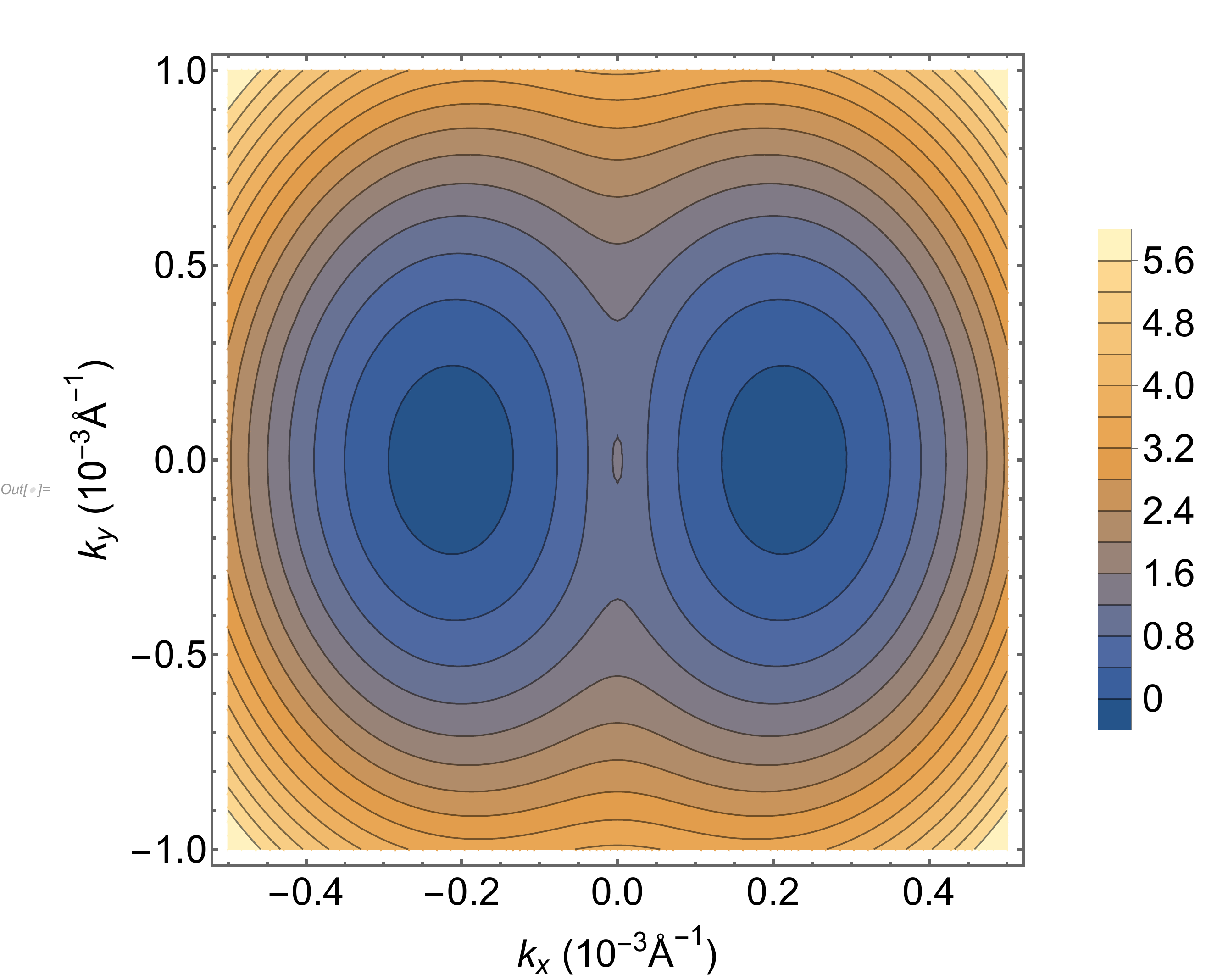}}
	\caption{(a) Structure of monolayer phosphorene. (b) Contour plot of the eigenenergy (in units of $\mu$eV) for the lower conduction band of monolayer phosphorene minus $\Delta$.} 
	\label{fig:bp}
\end{figure}

Monolayer phosphorene has a puckered honeycomb structure formed by phosphor atoms as depicted in Fig.~\ref{fig:bp} (a). The buckling reduces 3-fold symmetries in the honeycomb lattice to 2-fold, leading to the $D_{2h}$ symmetry and suggests its potential for Cartesian multipole moments with 2nd order in the spatial components. The low-energy Hamiltonian of multi-layer black phosphorous has been shown to be \cite{rudenko_2014, Ezawa_2014, rudenko_2015, fukuoka_2015, baik_2015, deSousa_2017,chaves_2017}
\begin{eqnarray}\label{eq:bpH}
	h_0(\mathbf k) = \left( \frac{\hbar^2 k_y^2}{2 m_y} + \Delta  \right) \sigma_x - \hbar v_x k_x \sigma_y
\end{eqnarray}
where $\boldsymbol{\sigma}$ and $\mathbf s$ are the Pauli matrix vectors in the pseudospin and real spin basis, respectively, and $\Delta$ is a gap of structural origin.

Because the pristine phosphorene has spatial inversion symmetry, all bands are doubly degenerate. We therefore consider a Rashba-type spin-orbit coupling that is induced by a perpendicular electric field. As a first approximation the form of the Rashba spin-orbit coupling can be obtained from the following term added to a single-orbital nearest-neighbor tight-binding model on the phosphorene lattice
\begin{eqnarray}\label{eq:HRTB}
	H_R = \sum_{\langle i \alpha, j\beta\rangle ab} \imath \lambda^R_{i\alpha,j\beta} \left(\hat{r}_{i\alpha,j\beta}\times \mathbf s_{ab}\right)\cdot \hat{z} c_{i\alpha a}^\dag c_{j\beta b} 
\end{eqnarray}
where $ij$ label unit cells, $\alpha\beta$ label sublattices, and $ab$ label spin. The Rashba spin-orbit coupling strength $\lambda^R$ only takes two possible values depending on whether the $(i\alpha,j\beta)$ nearest-neighbor bond is between same or different atomic planes of the phosphorene. The resulting Rashba term in the continuum model is (see Sec. III of \cite{supp})
\begin{eqnarray}
	h_R(\mathbf k) = \Lambda^R_0 \sigma_y s_y  + \Lambda_1^R k_y \sigma_x s_x - \Lambda^R_2 k_x\sigma_x s_y
\end{eqnarray}
where $\Lambda^R_{0,1,2}$ are constants depending on the lattice parameters as well as $\lambda^R$ in Eq.~\eqref{eq:HRTB}. Note that the first term was proposed in doped multilayer phosphorene \cite{baik_2015}, and that projecting $h_R$ to the eigenstate basis of $h_0$ leads to the familiar anisotropic Rashba spin-orbit coupling identified in \cite{popovic_2015,farzaneh_2019} through DFT calculations. However, the $k-$dependent terms are crucial for the spin octupole moments discussed below. We start our analysis by ignoring $\Lambda^R_2$ since it results in a similar but much smaller spin splitting in the eigenstate basis compared to that due to $\Lambda^R_0$ (see Sec.~III in \cite{supp}). In this case the eigenvalues of $h=h_0 +h_R$ can be simply obtained by squaring it twice:
\begin{eqnarray}
	\epsilon_{l,s} = l\sqrt{a^2 + b^2 + c^2 + d^2 + 2s \sqrt{a^2 c^2 + b^2d^2 + c^2 d^2 }}
\end{eqnarray}
where $a,b,c,d$ are the coefficients of $\sigma_x,\sigma_y,\sigma_xs_x,\sigma_ys_y$ in $h$, respectively, and $l,s=\pm 1$ label the four bands. One can easily show that two extrema of the bands are located at $\pm \mathbf K_v  = \pm \Lambda_0^R/(\hbar v_x)\hat{x}\approx \pm 2\times 10^{-4} {\rm \AA}^{-1}$, estimated using the DFT-extracted Rashba spin-orbit splitting under a 2.6 V/\AA\, electric field reported in \cite{popovic_2015}. This is consistent with Fig.~\ref{fig:bp} which shows a contour plot of the lower conduction band obtained by diagonalizing $h(\mathbf k)$ including $\Lambda_2^R$. 

We next focus on the center of each valley and consider the octupole component $\mathcal{M}_{xy}^k$ since such a component corresponds to a nontrivial staggered spin accumulation pattern at the corners of a rectangular sample of phosphorene (see below). Still ignoring $\Lambda_2^R$, when $k_y = 0$, $s_y = s = \pm 1$ is a good quantum number, and one can obtain the eigenstates of $h(\mathbf K_v)$ exactly:
\begin{eqnarray}\label{eq:psils}
	\psi_{l,-1}(\mathbf K_v) = \frac{1}{\sqrt{2}} \begin{pmatrix}
		1 \\
		l
	\end{pmatrix},\; \psi_{l,1}(\mathbf K_v) = \frac{1}{\sqrt{2}} \begin{pmatrix}
		1 \\
		l e^{i\theta_R}
	\end{pmatrix}
\end{eqnarray}
where $\theta_R\equiv 2\Lambda_0^R/\Delta $. The velocity operator at $\mathbf K_v$ is $\mathbf v =-  v_x \sigma_y \hat{x} +\frac{1}{\hbar}\Lambda_1^R \sigma_x s_x \hat{y}$.

For simplicity we consider the lower conduction band only ($l=1,s =1$ at $\mathbf K_v$, $l=1,s = -1$ at $-\mathbf K_v$). We found that in this case only $\mathcal{M}_{xy }^x\neq 0$ and the nonzero contribution comes from the following term
\begin{eqnarray}\label{eq:Mxyx}
	\mathcal{M}_{xy}^x &=&- \hbar^2 {\rm Re} \sum_{\substack{l \neq n \\ m \neq n}}\frac{s^x_{nm}(v^y_{ln}  v^x_{ml}  + v^x_{ln}  v^y_{ml} )}{(\epsilon_{n } - \epsilon_{l})(\epsilon_{n} - \epsilon_{m}) } \\\nonumber
	&\approx& -\frac{\hbar^2 v_x \Lambda^R_1 \Delta}{4(\Lambda_0^R)^3} \approx -1.66\times 10^6\, {\rm \AA}^2 \cdot \frac{\hbar}{2}
\end{eqnarray}
Such a large result originates from the tiny splitting between the two conduction bands at the valley center. It is also easy to check that at the opposite valley $\mathcal{M}_{xy}^x (-\mathbf K_v) \approx \frac{\hbar^2 v_x \Lambda^R_1 \Delta}{4(\Lambda_0^R)^3}$ as expected. 

\begin{figure}[ht]
	\centering
	\subfloat[]{\includegraphics[width=2.8 in]{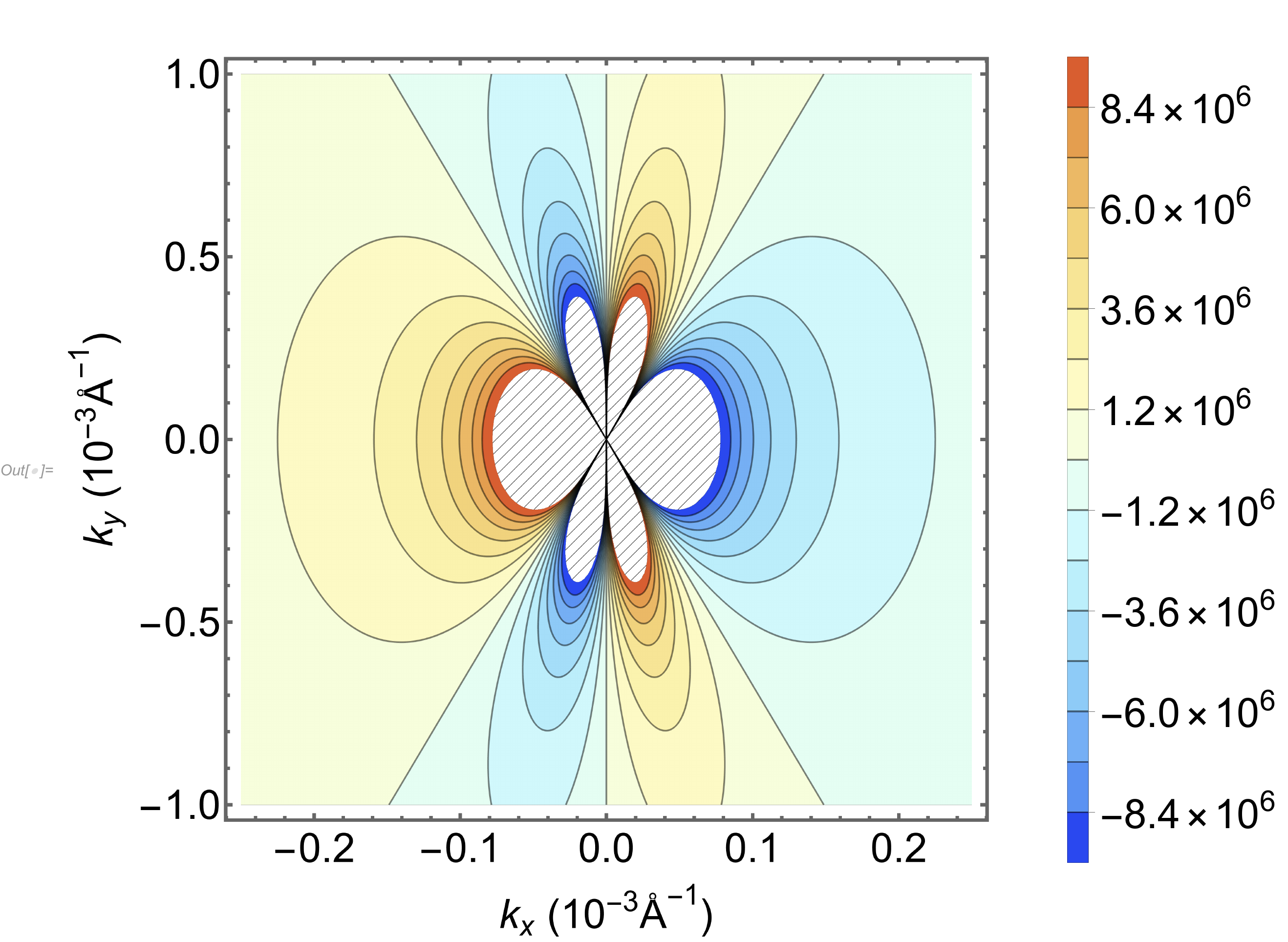}}\;\;\subfloat[]{\includegraphics[width=2.8 in]{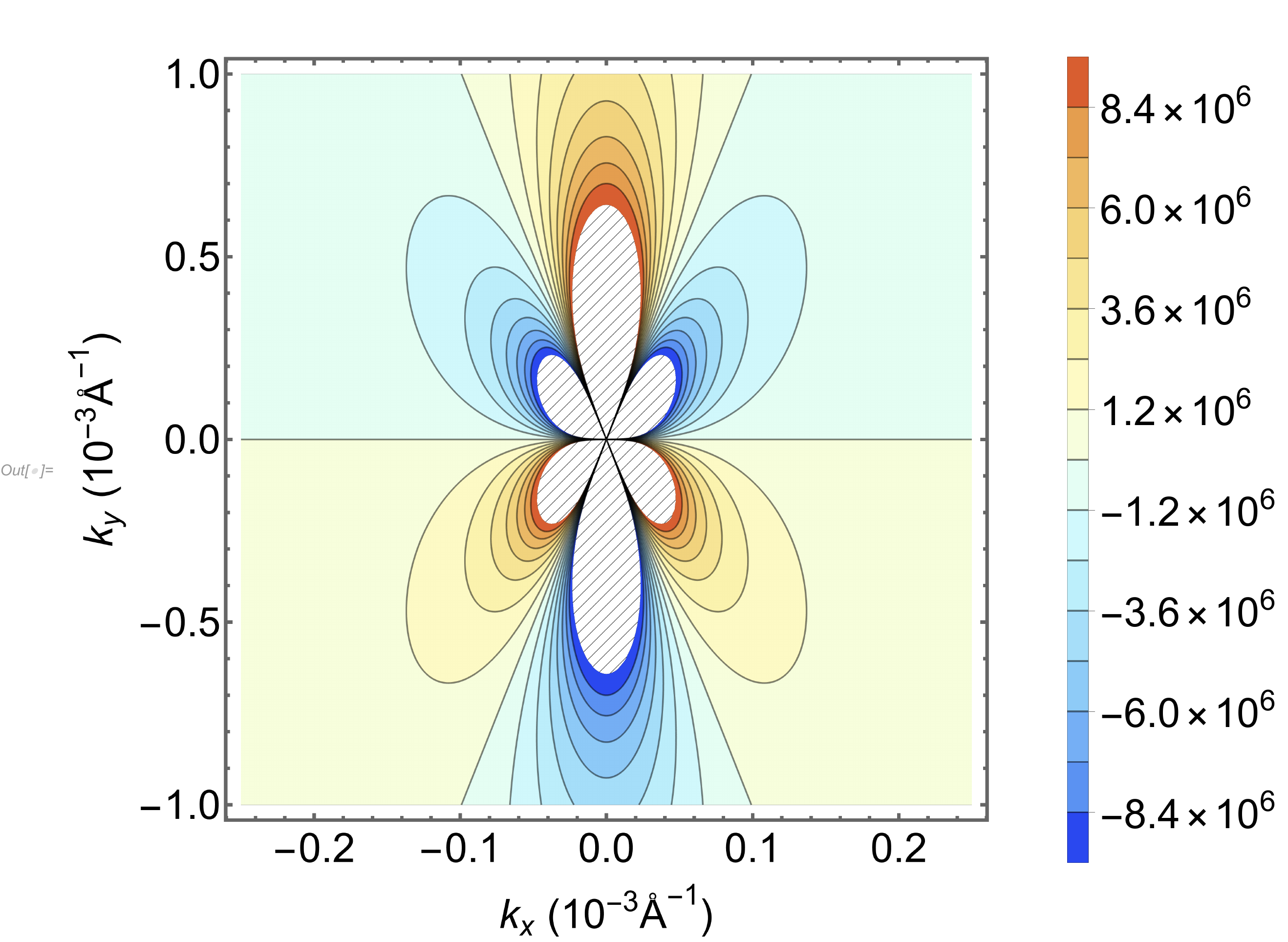}}
	\caption{(a) $\mathcal{M}_{xy}^x$ and (b) $\mathcal{M}_{xy}^y$ of the lowest conduction band of phosphorene. The regions where their values become too large due to the vanishing spin-orbit splitting between the two conduction bands are crossed out for better illustration.} 
	\label{fig:mxykphosphorene}
\end{figure}

Figure.~\ref{fig:mxykphosphorene} plots $\mathcal{M}_{xy}^x$ and $\mathcal{M}_{xy}^y$ for the lowest conduction band near the Brillouin zone center with $\Lambda_2^R$ included. One can see that the above analytic results are in agreement with that depicted in Fig.~\ref{fig:mxykphosphorene} (a). The vanishing $\mathcal{M}_{xy}^y(\pm \mathbf K_v)$ is also confirmed by Fig.~\ref{fig:mxykphosphorene} (b). The large size of $\mathcal{M}_{xy}^{x,y}$ is again due to the tiny spin-orbit splitting between the two conduction bands. The net multipole moments induced by, e.g., an electric field therefore involve strong cancellation between contributions of opposite signs from the two conduction bands at the Fermi energy, which we discuss in more detail in the next section.

\section{Current-induced magnetic multipole moments in phosphorene}\label{sec:current}
Since the spin multipole moments defined in this work are properties of individual quasiparticles, their creation in real space by external perturbations can be easily formulated in terms of quasiparticle excitations, i.e. intra-band transitions, induced by the perturbations. The intra-band contributions are also expected to be dominant in clean systems with a long relaxation time. Possible issues of considering inter-band effects will be discussed at the end of this paper. In this section we present a Boltzmann theory for the creation of magnetic octupole moments by electric currents in phosphorene.

\begin{figure}[ht]
	\centering
	\includegraphics[width=2.9 in]{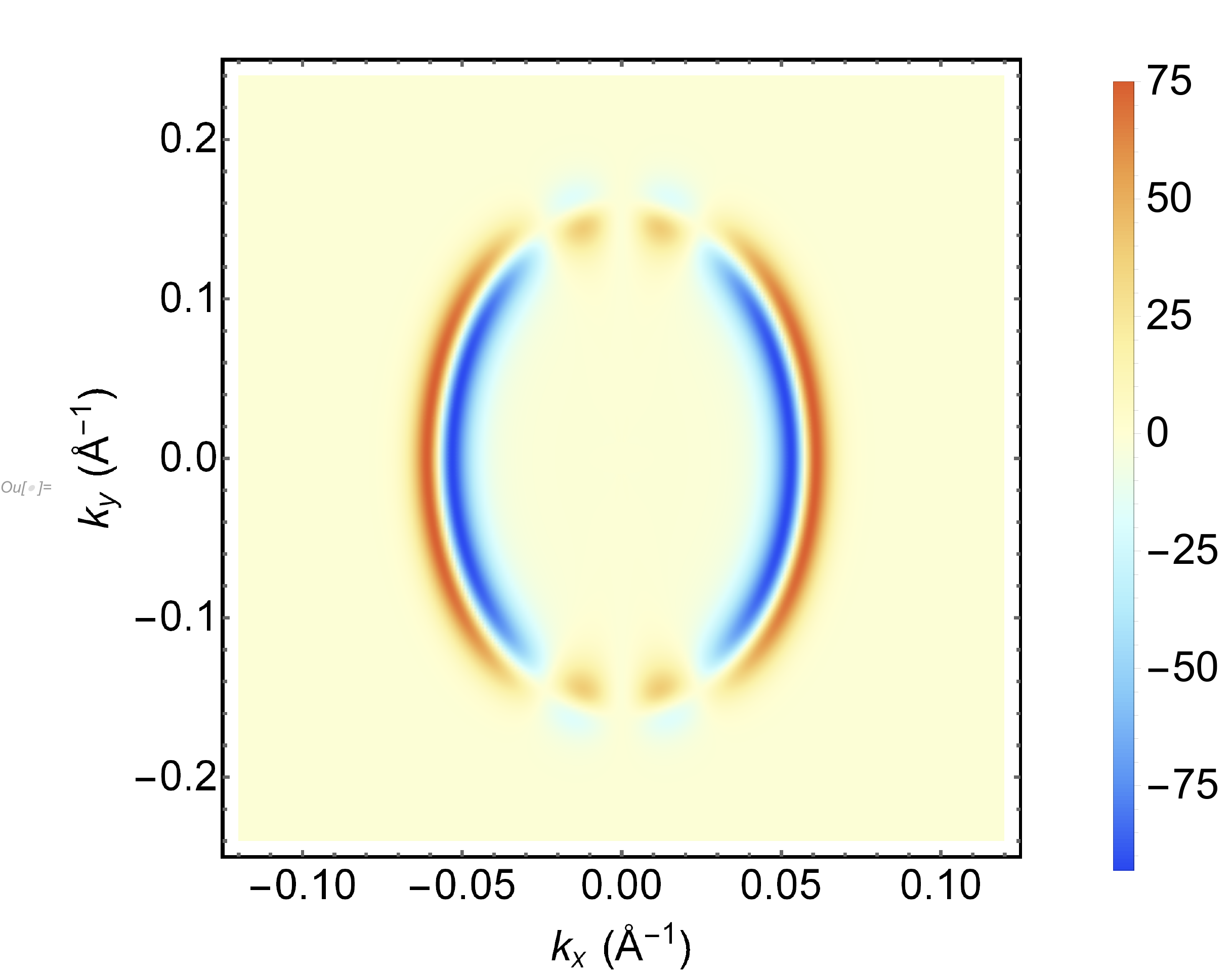}
	\caption{Integrand of Eq.~\eqref{eq:cimxyx} ($k=x$) with $E_F-\Delta = 0.1$ eV, $k_B T = 0.01$ eV, and $\mathbf E\parallel \hat{x}$.} 
	\label{fig:cimxyx}
\end{figure}

The nonequilibrium distribution function that is first order in the electric field $\mathbf E$ solved from the Boltzmann equation with a constant relaxation time $\tau$ is $f^{(1)} = e\tau\frac{\partial f_0}{\partial\epsilon_{n\mathbf k}}\mathbf v_{n\mathbf k}\cdot \mathbf E$. Therefore the nonequilibrium octupole moment density is
\begin{eqnarray}\label{eq:cimxyx}
	\langle \mathcal{M}_{xy}^k \rangle = e\tau \sum_{n}\int \frac{d^2\mathbf k}{(2\pi)^2} \mathcal{M}_{xy}^k \frac{\partial f_0}{\partial\epsilon_{n\mathbf k}}\mathbf v_{n\mathbf k}\cdot \mathbf E
\end{eqnarray}
which can be generalized to arbitrary multipole moments. The integrand of the above equation ($k=x$) for $E_F-\Delta = 0.1$ eV, $k_B T = 0.01$ eV, and $\mathbf E\parallel \hat{x}$ is shown in Fig.~\ref{fig:cimxyx}. One can clearly see that the contribution due to each conduction band has a nonzero Fermi surface integral, although the two bands have opposite contributions which largely cancel with each other. Taking $\hbar/\tau = 0.1$ eV, the result of Eq.~\eqref{eq:cimxyx} per electric field in the units of $\rm V/{\rm \AA}$ is $-1.65\times 10^{-2}$ $\rm \mu_B$. 

To understand the implication of this result, we note that in a macroscopic phosphorene sample of size $L_x\times L_y$, an octupole moment $\mathcal{M}_{xy}^x$ corresponds to corner spin with the size $S_c$ and alternating signs at the four corners. Namely, $\mathcal{M}_{xy}^x = 4\times \frac{L_x L_y}{4} S_c$, or equivalently the density of the octupole moment is equal to the size of the corner spin. The above result indicates that with a Rashba spin-orbit coupling induced by a 2.6 V/$\rm \AA$ perpendicular electric field, the corner-spin-electric-field response has a rough size of $-1.65\times 10^{-2}\,\rm \mu_B V^{-1} \AA$. The other components of the octupole-electric-field response function can be similarly obtained and can be understood as different corner spin responses as well if the two spatial indices are different. For comparison, using a formula similar to Eq.~\eqref{eq:cimxyx} one can calculate the current-induced uniform spin density due to the Rashba spin-orbit coupling. The same set of parameters give a $y$-spin density of $-4.8\times 10^{-4}\,\rm \mu_B/\AA^2$ per $1\,\rm V/\AA$ electric field along $x$. Therefore with the present parameter values the uniform current-induced spin density in an area of $\sim 1\,{\rm nm}^2$ near a corner may overshadow that due to the corner spin, although they are polarized in different directions. Such a fact may need to be considered when designing experiments for measuring the corner spin. 

Finally, we point out that a current (along $x$) carrying the octupole moment $\mathcal{M}_{xy}^x$ can be created by a nonlinear response to the electric field (along $x$). Note that such a current does not arise from linear response since it is forbidden by symmetry, which can also be seen from Fig.~\ref{fig:cimxyx}. The nonequilibrium distribution function of the second order in $\mathbf E$ is
\begin{eqnarray}
f^{(2)} = e^2\tau^2 E_\alpha E_\beta \left( \frac{\partial^2 f_0}{\partial \epsilon_{n\mathbf k }^2} v_{n\mathbf k}^\alpha v_{n\mathbf k}^\beta +\frac{1}{\hbar^2}\frac{\partial f_0}{\partial \epsilon_{n \mathbf k}} \frac{\partial^2\epsilon_{n\mathbf k}}{\partial k_\alpha\partial k_\beta}\right).
\end{eqnarray}
where the inverse effective mass tensor in the second term is (Eq.~(48) in \cite{supp})
\begin{eqnarray}
\frac{1}{\hbar^2}\partial_{\alpha}\partial_{\beta} \epsilon_{n\mathbf k} =\frac{1}{\hbar^2} (\partial_\alpha \partial_\beta H_{\mathbf k})_{nn} + 2{\rm Re} \sum_{m\neq n} \frac{v^\alpha_{nm,\mathbf k} v^\beta_{mn,\mathbf k}}{\epsilon_{n\mathbf k} - \epsilon_{m\mathbf k}}.
\end{eqnarray}
When $\mathbf E\parallel \hat{x}$, $f^{(2)}$ is symmetric with respect to $k_x = 0$, so is $\mathcal{M}_{xy}^xv_{nk}^x$. Therefore the octupole current $\langle \mathcal{M}_{xy}^x v_x \rangle \neq 0$. Such a conclusion can also be reached by considering the symmetry of the phosphorene under a perpendicular electric field (Sec.~IV in \cite{supp}). In general, if a multipole moment vanishing in equilibrium becomes nonzero through linear response to an electric field, a second-order response to the same electric field can drive a current of that multipole moment.

\section{Discussion}\label{sec:discussion}
Our work shows that gauge dependence and origin dependence can be separate issues in studying multipole moments of solid state materials. On the one hand, classical multipole moments in general depend on origin. On the other hand, explicitly fixing the origin does not necessarily fix the gauge for a quantum multipole moment, since the former only sets a linear-in-$\mathbf k$ phase of a U(1) gauge transformation. For the case of spin magnetic multipole moments, although subtracting $\mathbf r_c$ from $\mathbf r$ can lead to gauge-invariant formulas for $l\le3$, the procedure in general fails for $l>3$. 

Our work only discussed the intrinsic multipole moments of Bloch wave packets that do not depend on the shape of the wave packets. Macroscopic multipole moments of a given system defined in a thermodynamic manner in general involves more contributions. However, inspired by the studies of orbital magnetization and spin toroidization, it may be possible to express these other contributions as boundary terms due to lower-order wave-packet multipole moments, an interesting question awaiting future investigation. Similarly, response functions of mutlipole moments to external perturbations can also have contributions due to the off-diagonal part of the nonequilibrium density matrix. These contributions are, however, not as well defined since the off-diagonal matrix elements of certain multipole moment operator are needed, which are not uniquely defined in an infinite crystal \cite{chen2019spin}. A more sensible view is to only consider the response functions that can be directly measured. For example the current-induced octupole moment in phosphorene is essentially the response of staggered corner spins to electric fields. The currents of multipole moments, similar to spin currents, are strictly speaking a conceptual tool rather than a well-defined observable.

For prototypical materials hosting magnetic multipole moments, although we only discussed phosphorene, multi-layer black phosphorus should also work since it has a similar low-energy Hamiltonian as the former \cite{deSousa_2017}. In three dimensions it will also be interesting to look for hexadecapole moments with $xyz$ spatial indices, since they would lead to staggered corner spins in a cubic sample. We expect orthorhombic materials with either time-reversal or inversion symmetry broken to be promising candidates for this effect. 

\begin{acknowledgements} 
MT and HC were supported by NSF CAREER grant DMR-1945023. The authors are grateful to Qian Niu and Yang Gao for valuable discussions.
\end{acknowledgements} 

\appendix
\section*{Competing interests}
The authors declare that they have no competing interests.

\section*{Data and code availability}
All data and codes presented in the paper, if not present in the main text or supplementary information, are available upon reasonable request from the corresponding author.

\bibliography{cim3_ref}

\end{document}